\documentclass[structabstract]{aa}
\usepackage{graphicx}
\usepackage{natbib}
\bibpunct{(}{)}{;}{a}{}{,} 

\begin{document}
   \title{The standard flare model in three dimensions}

   \subtitle{II. Upper limit on solar flare energy}

   \author{G. Aulanier  \inst{1}
          \and
          P. D\'emoulin \inst{1}
          \and
          C.~J. Schrijver \inst{2}
          \and
          M. Janvier    \inst{1}
          \and
          E. Pariat     \inst{1}
          \and
          B. Schmieder  \inst{1}
          }

   \institute{LESIA, Observatoire de Paris, CNRS, UPMC, Univ. Paris 
              Diderot, 5 place Jules Janssen, 92190 Meudon\\
              \email{guillaume.aulanier@obspm.fr}
         \and
             Lockheed Martin Advanced Technology Center, 3251 Hanover 
             Street, Palo Alto, CA 94304, USA\\
             }

   \date{Received 19 September 2012 ; Accepted 14 November 2012}

 
  \abstract
   {Solar flares strongly affect the Sun's atmosphere as well as the 
   Earth's environment. Quantifying the maximum possible energy of solar 
   flares of the present-day Sun, if any, is thus a key question in 
   heliophysics.}
   {The largest solar flares observed over the past few decades have reached 
   energies of a few times $10^{32}$ ergs, possibly up to $10^{33}$ ergs. Flares 
   in active Sun-like stars reach up to about $10^{36}$ ergs. In the absence 
   of direct observations of solar flares within this range, complementary 
   methods of investigation are needed to assess the probability of solar 
   flares beyond those in the observational record.} 
    {Using historical reports for sunspot and solar active region 
    properties in the photosphere, we scaled to observed solar values 
    a realistic dimensionless 3D MHD simulation for eruptive flares, 
    which originate from a highly sheared bipole. This enabled us to 
    calculate the magnetic fluxes and flare energies in the model 
    in a wide paramater space.}
   {Firstly, commonly observed solar conditions lead to modeled magnetic 
   fluxes and flare energies that are comparable to those estimated from 
   observations. Secondly, we evaluate from observations that $30\%$ of the area 
   of sunspot groups are typically involved in flares. This is related to the 
   strong fragmentation of these groups, which naturally results from sub-photospheric 
   convection. When the model is scaled to $30\%$ of the area of the largest 
   sunspot group ever reported, with its peak magnetic field being set to 
   the strongest value ever measured in a sunspot, it produces a flare with 
   a maximum energy of $\sim6\times10^{33}$ ergs.} 
   {The results of the model suggest that the Sun is able to produce flares up 
   to about six times as energetic in total solar irradiance (TSI) fluence 
   as the strongest directly-observed 
   flare from Nov 4, 2003. Sunspot groups larger than historically reported would 
   yield superflares for spot pairs that would exceed tens of degrees in 
   extent. 
   We thus conjecture that superflare-productive 
   Sun-like stars should have a much stronger dynamo than in the Sun.}


   \maketitle


\section{Introduction}
\label{secintro}
   
Solar flares result from the abrupt release of free magnetic energy, that 
has previously been stored in the coronal magnetic field by flux emergence 
and surface motions \citep{Forbes06}. Most of the strongest flares are 
eruptive \citep[as reviewed by][]{SchrijverCospar09}. For the latter, the 
standard model attributes the flare energy release to magnetic reconnection 
that occurs in the wake of coronal mass ejections 
\citep{Shibata95,LinForbes00,Moore01,Priest02}. 

Several flare-related phenomena impact the solar atmosphere itself. To be 
specific, there are photospheric sunquakes 
\citep{Zhar11}, 
chromospheric ribbons 
\citep{Sch87}, 
coronal loop restructuration 
\citep{Warren11} 
and oscillation 
\citep{Nak99}, 
large-scale coronal propagation fronts 
\citep{Dela08}, 
and driving of sympathetic eruptions 
\citep{SchrijverTitle11}. 
In addition to solar effects, flare-related irradiance enhancements 
\citep{Woods04}, 
solar energetic particles 
\citep[SEPs, ][]{MassonKlein09} 
and coronal mass ejections
\citep[CMEs, ][]{Vour10} 
constitute major drivers for space weather, and are responsible for various 
environmental hazards at Earth 
\citep{Schwenn06,Pulk07}. 

For all these reasons, it would be desireable to know whether or 
not there is a maximum for solar flare energies, and if so, what its 
value is. 

On the one hand, detailed analyses of modern data from the past half-century 
imply that solar flare energies range from $10^{28}$ to $10^{33}$ ergs, 
with a power-law distribution that drops above $10^{32}$ ergs 
\citep{Schrij12}. The maximum value there corresponds to an estimate 
for the strongest directly-observed flare from Nov 4, 
2003. 
Saturated 
soft X-ray observations showed that this flare was above the X28 
class, and model interpretations of radio observations of Earth's ionosphere 
suggested that it was X40 \citep{Brod05}. Due to the limited range in time 
of these observations, it is unclear whether or not the Sun has been 
-or will be- able to produce more energetic events. For example, the 
energy content of the first-ever observed solar flare on Sept 1, 1859 
\citep{Car59,Hod59} has been thoroughly debated 
\citep{Mc01,Tsu03,Cliv04,Wolff12}. 
On the other hand, precise measurements on unresolved active Sun-like stars 
have revealed the existence of so-called superflares, even in slowly-rotating 
and isolated stars \citep{Schae00,Mae12}. Their energies have been 
estimated to be between a few $10^{33}$ ergs to more than $10^{36}$ ergs. 
Unfortunately, it is still unclear whether or not the Sun can 
produce such superflares, among other reasons because of the lack 
of reliable information on the starspot properties of such stars 
%
\citep{Ber05,Stras09}.
%

So as to estimate flare energies, a method complementary to observing solar 
and stellar flares is to use solar flare models, and to constrain the parameters 
using observational properties of active regions, rather than those of the flares themselves. 
In the present paper, we perform such an analysis. Since analytical approaches are 
typically oversimplified for such a purpose, numerical models are likely to 
be required. Moreover, incorporating observational constraints not only precludes 
the use of 2D models, but also restrict the choice to models that 
have already proven to match various solar observations to some acceptable degree. 

We use a zero-$\beta$ MHD simulation of an eruptive flare \citep{Aula10,Aula12} 
that extends the standard flare model in 3D. Dedicated analyses of the 
simulation, as recalled hereafter, have shown that this model successfully reproduced 
the time-evolution and morphological properties of active region magnetic fields 
after their early emergence stage, of coronal sigmoids from their birth to their 
eruption, of spreading chromospheric ribbons and sheared flare loops, of tear-drop 
shaped CMEs, and of large-scale coronal propagation fronts. We scaled the model 
to solar observed values as follows: we incorporate observational constraints known from 
previously reported statistical studies regarding the magnetic flux of active regions, 
as well as the area and magnetic field strength of sunspot groups. This method allows 
one to identify the maximum flare energy for realistic but extreme solar conditions, 
and to predict the size of giant starspot pairs that are required to produce superflares. 


\section{The eruptive flare model}
\label{sec1}

\subsection{Summary of the non-dimensionalized model}

The eruptive flare model was calculated numerically, using the {\em observationally driven 
high-order scheme magnetohydrodynamic} code \citep[OHM:][]{Aula05a}. The calculation 
was performed in the pressureless resistive MHD approximation, using non-dimensionalized 
units, in a $251 \times 251 \times 231$ non-uniform cartesian mesh. Its uniform resistivity 
resulted in a Reynolds number of about $R_m\sim10^3$. The simulation settings 
are thoroughly described in \citet{Aula10,Aula12}. 

In the model, the flare resulted from magnetic reconnection occuring at a nearly 
vertical current sheet, gradually developing in the wake of a coronal mass ejection. 
The reconnection led to the formation of ribbons and flare loops \citep{Aula12}. 
The CME itself was triggered by the ideal loss-of-equilibrium 
of a weakly twisted coronal flux rope \citep{Aula10}, corresponding to the torus 
instability \citep{KliTor06,DemAula10}. During the eruption, a coronal propagation 
front developed at the edges of the expanding sheared arcades surrounding the flux 
rope \citep{Schrijver11}. Before it erupted, the flux rope and a surrounding sigmoid 
were progressively formed in the corona \citep{Aula10,Sav12}, above a slowly shearing 
and diffusing photospheric bipolar magnetic field. This pre-eruptive evolution was 
similar to that applied in past symmetric models \citep{vanBalle89,Amari03b}, and 
they matched observations and simulations for active regions during their late 
flux emergence stage and their subsequent decay phase \citep[e.g.][]{vanDriel03,Arch04,Green11}. 

The magnetic field geometry of the modeled eruptive flare is shown in Fig.~\ref{fig1}. 
The left panel clearly shows the asymmetry of the model. A $27\%$ flux imbalance in 
the photosphere, in favor of the positive polarity, manifests itself as open magnetic 
field lines rooted in the positive polarity, at the side of the eruption. This 
asymmetry was set in the model so as to reproduce typical solar active regions, with 
a stronger (resp. weaker) leading (resp. trailing) polarity. In the 
right panels, the field of view corresponds to the size of the magnetic bipole 
$L^{\mbox{\rm{\tiny bipole}}}$, as used for physical scaling hereafter. 

If one assumes a sunspot field of $B_z^{\mbox{\rm{\tiny max}}}=3500$ G, then the 
isocontours that cover the widest areas correspond to $B_z^{\mbox{\rm{\tiny max}}}/5
=\pm 700$ G. Since this is the minimum magnetic field value for sunspot penumbrae 
\citep{Solanki06}, those isocontours correspond to the outer edge of the modeled 
sunspots. With these settings, the total sunspot area in the model is about 
half of the area of the field of view being shown in Fig.~\ref{fig1}, {\em right}. 
So with $B_z^{\mbox{\rm{\tiny max}}}=3500$ G the sunspot area is $f^{-1}\, 
(L^{\mbox{\rm{\tiny bipole}}})^2$, with $f\sim2$, while a lower value 
for $B_z^{\mbox{\rm{\tiny max}}}$ implies a higher value for $f$. 

During the pre-eruptive energy storage phase, the combined effects of shearing motions 
and magnetic field diffusion in the photosphere eventually resulted in the development 
of magnetic shear along the polarity inversion line, over a length of about 
$L^{\mbox{\rm{\tiny bipole}}}$. This long length presumably results in the modeled 
flare energy to be close to its maximum possible value, given the distribution of 
photospheric flux \citep{Fal08,Moore12}. 

\subsection{Physical scalings}

     \begin{figure*}
     \sidecaption
     \includegraphics[width=12cm]{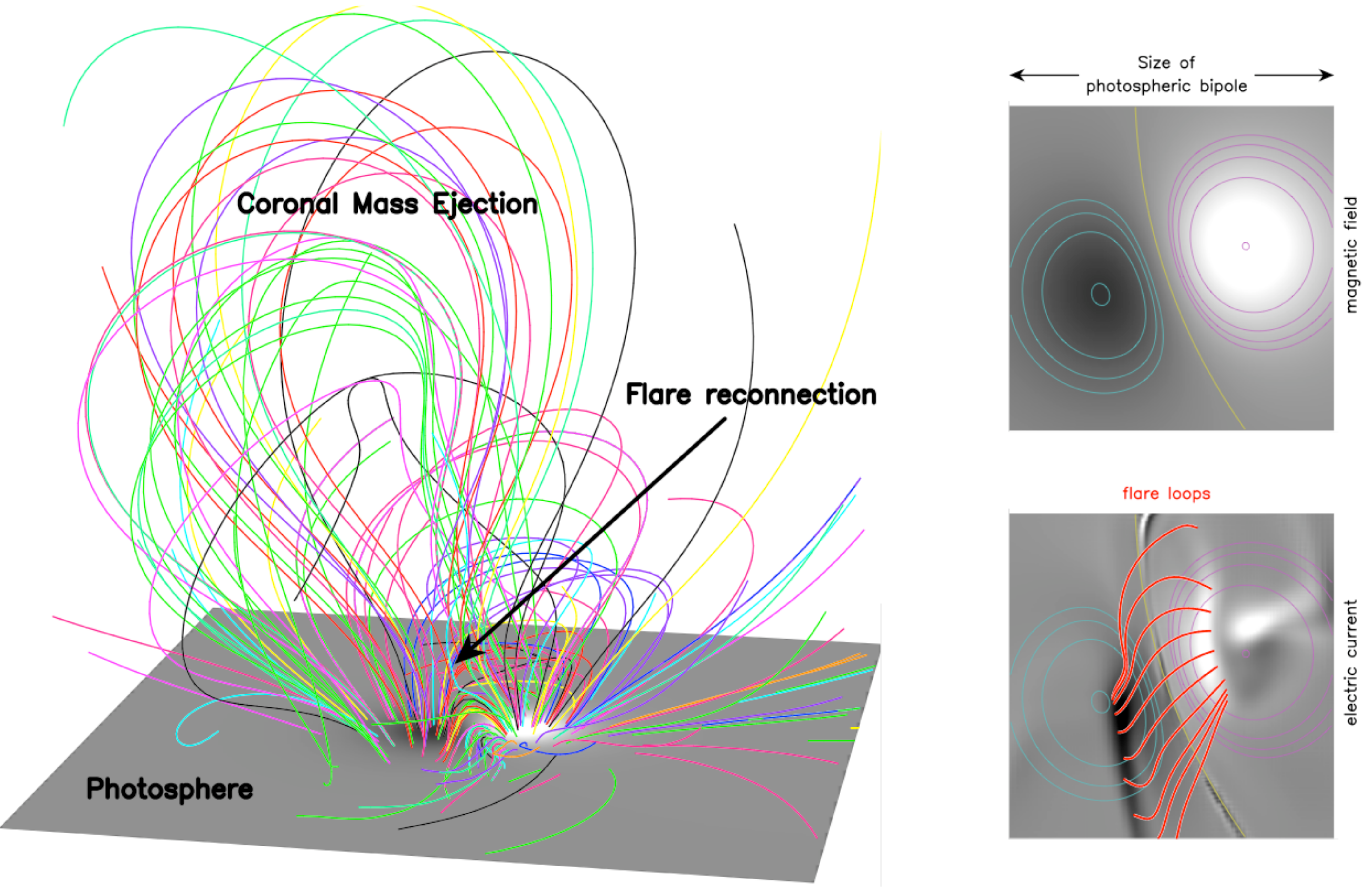}
     \caption{
Eruptive flare model. 
{\em [left:]} Projection view of randomly plotted coronal magnetic field 
lines. The grayscale corresponds to the vertical component of the 
photospheric magnetic field $B_z$. 
{\em [right:]} Photospheric bipole viewed from above. 
The pink (resp. cyan) isocontours stand for positive (resp. 
negative) values of $B_z^{\mbox{\rm{\tiny max}}}/1.1, 2, 3, 4, 5$. 
The yellow isocontour shows the polarity inversion line $B_z=0$. 
{\em [right-top:]} The grayscale for $B_z$ is the same as in the left panel. 
{\em [right-bottom:]} The grayscale shows the vertical component of the 
photospheric electric 
%
currents. Strong elongated white/black patches highlight flare ribbons. 
The red lines show representative post-reconnection flare loops, rooted 
in the flare ribbons. 
%
             }
     \label{fig1}
     \end{figure*}

The MHD model was calculated in a wide numerical domain of size $20\times20\times30$, 
with a magnetic permeability $\mu=1$, using dimensionless values $B_z^{\mbox{\rm{\tiny 
max}}}=8$ in the dominant polarity, and $L^{\mbox{\rm{\tiny bipole}}}=5$. These 
settings resulted in a dimensionless photospheric flux inside the dominant polarity 
of $\phi=42$ \citep{Aula10}, and a total pre-eruptive magnetic energy of 
$E^{\mbox{\rm{\tiny bipole}}}=225$. 

Throughout the simulation, a magnetic energy of $E^{\mbox{\rm{\tiny model}}}
=19\%\, E^{\mbox{\rm{\tiny bipole}}}=42$ was released. Only $5\%$ of this 
amount was converted into the kinetic energy of the CME. These numbers have 
been presented and discussed in \citet{Aula12}. The remaining $95\%\, 
E^{\mbox{\rm{\tiny model}}}$ of the magnetic energy release can then be 
attributed to the flare energy itself. 

It must be pointed out that the simulation did not cover the full duration 
of the eruption. Indeed, numerical instabilities eventually prevented us from 
pursuing it with acceptable diffusion coefficients. Nevertheless, the rate 
of magnetic energy decrease had started to drop before the end of the 
simulation, and the electric currents within the last reconnecting field 
lines where relatively weak. On the one hand, this means that the total 
energy release $E^{\mbox{\rm{\tiny model}}}$ is expected to be slightly 
higher, but presumably not by much. On the other hand, 
the relatively low $R_m$ value of the simulation implies that some 
amount of $E^{\mbox{\rm{\tiny model}}}$ should be attributed to large-scale 
diffusion, rather than to the flare reconnection. 

Because of these numerical concerns, we consider thereafter that the flare 
energy in the model was about $E=40$, but this number should not be taken 
as being precise. Also, within the pressureless MHD framework of the 
simulation, the model cannot address which part of this energy is converted 
into heating, and which remaining part results in particle acceleration. 

\begin{figure*}
     \centering
     \includegraphics[width=\textwidth]{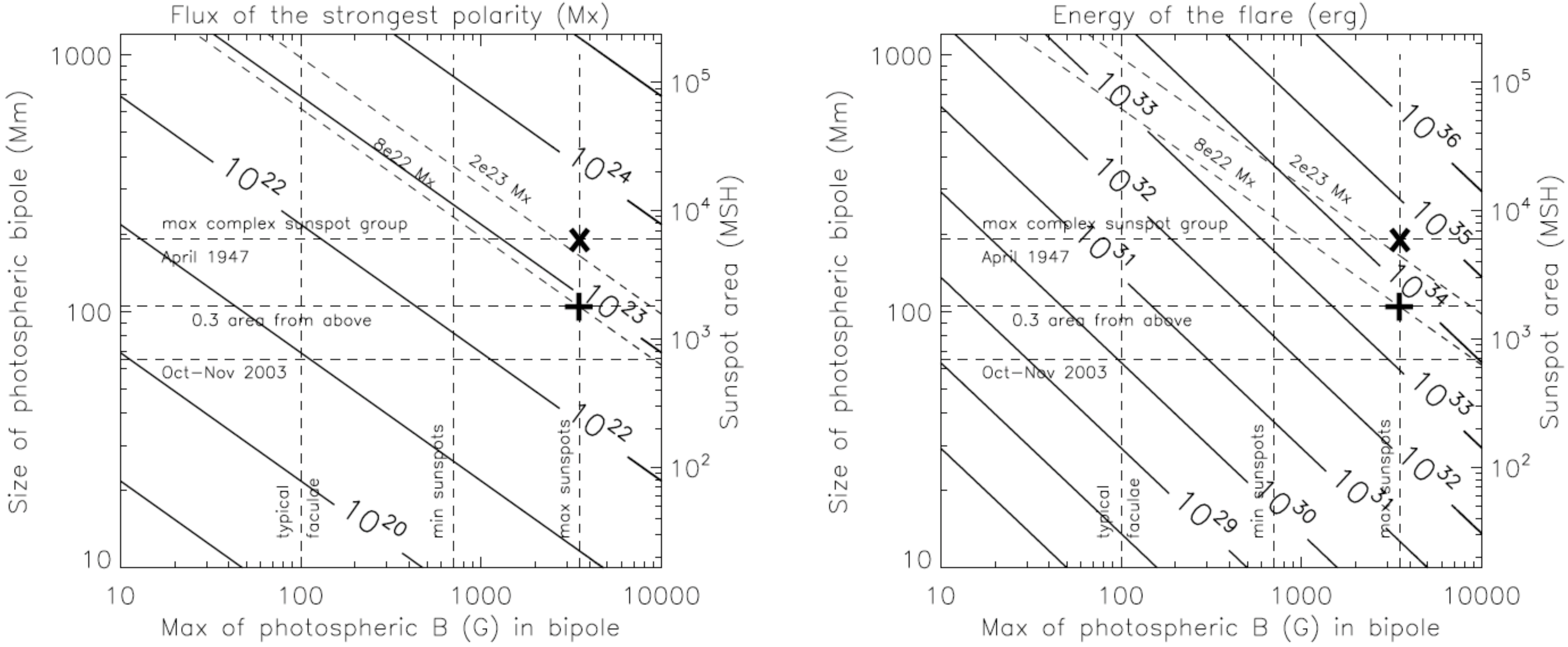}
     \caption{
Magnetic flux in the dominant polarity of the bipole, and magnetic 
energy released during the flare, calculated as a function of the 
maximum magnetic field and the size of the photospheric bipole. The 
$\times$ and $+$ signs correspond to extreme solar values. The 
former is unrealistic and the latter must be very rare (see text 
for details).
          }
     \label{fig2}
     \end{figure*}

It is straightforward to scale the model numbers given above into 
physical units. In the international system of units (SI), $\mu=4\pi10^{-7}$, 
the total magnetic flux $\phi$ and the total flare energy $E$ can then 
be written as 
\begin{eqnarray}
\label{eqflux1}
\phi &=& 42\, 
         \Bigg{(}\frac{B_z^{\mbox{\rm{\tiny max}}}}{8~{\mbox{\rm T}}}\Bigg{)} \, 
         \Bigg{(}\frac{L^{\mbox{\rm{\tiny bipole}}}}{5~{\mbox{\rm m}}}\Bigg{)}^2
         \, {\mbox{\rm Wb}}\, , \\
\label{eqenergy1}
E    &=& \frac{40}{\mu}\, 
         \Bigg{(}\frac{B_z^{\mbox{\rm{\tiny max}}}}{8~{\mbox{\rm T}}}\Bigg{)}^2 \, 
         \Bigg{(}\frac{L^{\mbox{\rm{\tiny bipole}}}}{5~{\mbox{\rm m}}}\Bigg{)}^3
         \, {\mbox{\rm J}}\, .
\end{eqnarray}
Rearranging these equations into commonly used solar units leads to:
\begin{eqnarray}
\label{eqflux}
\phi &=& 0.52\times10^{22}\, 
         \Bigg{(}\frac{B_z^{\mbox{\rm{\tiny max}}}}{{10^3~\mbox{\rm G}}}\Bigg{)} \, 
         \Bigg{(}\frac{L^{\mbox{\rm{\tiny bipole}}}}{50~{\mbox{\rm Mm}}}\Bigg{)}^2
         \, {\mbox{\rm Mx}}\, , \\
\label{eqenergy}
E    &=& 0.5\times10^{32}\, 
         \Bigg{(}\frac{B_z^{\mbox{\rm{\tiny max}}}}{{10^3~\mbox{\rm G}}}\Bigg{)}^2 \, 
         \Bigg{(}\frac{L^{\mbox{\rm{\tiny bipole}}}}{50~{\mbox{\rm Mm}}}\Bigg{)}^3
         \, {\mbox{\rm erg}}\, .
\end{eqnarray}
While the power-law dependences in these equations come from the definitions 
of flux and energy, the numbers themselves directly result from the MHD simulation, 
and not from simple order of magnitude estimates. 
So Eqs.~(\ref{eqflux}) and (\ref{eqenergy}) enable us to calculate the model 
predictions for a wide range of photospheric magnetic fields and bipole sizes. 
The results are plotted in Fig.~\ref{fig2}. In this figure, the right vertical 
axis is the total sunspot area within the model, being given by $f^{-1}\, 
(L^{\mbox{\rm{\tiny bipole}}})^2$ using $f=2$. It is expressed in micro solar 
hemispheres \citep[hereafter written MSH as in][although other notations can 
be found in the literature]{Baumann05}. Hereafter all calculated energies 
(resp. fluxes) will almost always be given in multiples of $10^{32}$ ergs 
(resp. $10^{22}$ Mx), for easier comparison between different values. 

Typical decaying active regions with $L^{\mbox{\rm{\tiny bipole}}}=200$ Mm, which 
contain faculae of $B_z^{\mbox{\rm{\tiny max}}}=100$ G, have $\phi=0.8\times10^{22}$ 
Mx and can produce moderate flares of $E=0.3\times10^{32}$ ergs. Also, $\delta$-spots 
with $L^{\mbox{\rm{\tiny bipole}}}=40$ Mm and $B_z^{\mbox{\rm{\tiny max}}}=1500$ G 
have a lower magnetic flux $\phi=0.5\times10^{22}$ Mx, but can produce twice stronger 
flares, with $E=0.6\times10^{32}$ ergs. These energies for typical solar active 
regions are in good agreement with those estimated from the total solar irradiance 
(TSI) fluence of several observed flares \citep{Kret11}. 

Other parameters can result in more or less energetic events. For example one 
can scale the model to the sunspot group from which the 2003 Halloween flares 
originated. Firstly, one can overplot our Fig.~\ref{fig1}, {\em right}, onto 
the center of the Fig.~2 in \citet{Schrijver06} and thus find an approximated 
size of the main bipole which is involved in the flare, out of the whole 
sunspot group. This gives a bipole size of the order of $L^{\mbox{\rm{\tiny 
bipole}}} \sim 65$ Mm. Secondly, observational records lead to a peak sunspot 
magnetic field of $B_z^{\mbox{\rm{\tiny max}}}=3500$ G \citep{Living12}. These 
scalings lead to $\phi=3\times10^{22}$ Mx and $E=13\times10^{32}$ ergs. 
The modeled $\phi$ is about one third of the flux of the dominant polarity 
as measured in the whole active region \citep{Kaza10}. Comparing this modeled 
flare energy $E$ with that of extreme solar flares that originated from this 
same active region, we find that it is twice as strong as that of the Oct 28, 
2003 X17 flare \citep{Schrij12}, and about the same as that of the Nov 4, 
2003 X28-40 flare, as can be estimated from \citet{Kret11} and \citet[]
[Eq.~1]{Schrij12}.


\section{Finding the upper limit on flare energy}
\label{sec3}

\subsection{Excluding unobserved regions in the parameter space}

We indicate in Fig.~\ref{fig2} the minimum and maximum sunspot magnetic fields 
as measured from spectro-polarimetric observations since 1957. They are respectively 
$700$ G in the penumbra, and $3500$ G in the umbra \citep{Solanki06,Pevtsov11}. 
The latter value is an extreme that has rarely been reported in sunspot 
observations, and it typically is observed in association with intense  
flaring activity \citep{Living12}. 

We also indicate the maximum area of sunspot groups, including both the 
umbras and the penumbras. They were measured from 1874 to 1976 \citep{Baumann05} 
and from 1977 to 2007 \citep{Hat08}. These sizes follow a log-normal 
distribution up $3000$ MSH, but there are a few larger groups. The 
largest one was observed in April 1947, and its area was about $5400-6000$ 
MSH \citep{Nic48,Taylor89}. 
%
%
For illustration, we provide in Fig.~\ref{fig3} one image of 
this sunspot group and one of its surrounding faculae and 
filaments, as observed with the Meudon spectroheliograph. 
%
%
Interestingly, this sunspot group did not generate strong geomagnetic 
disturbances. This could either be due to a lack of strong enough magnetic 
shear in the filaments which were located between the sunspots, or to the lack 
of Earth-directed CMEs that could have been launched from this region. 
However, several other large sunspot groups, whose areas were at least 
$3500$ MSH, did generate major geomagnetic storms. Among those are the 
March 1989 event, which led to the Quebec blackout \citep{Taylor89}, 
and the December 1128 event, which produced aurorae in Asia and which 
corresponds to the first reported sunspot drawing \citep{Willis01}. 
Therefore, we conservatively keep $6000$ MSH as the maximum value. 
%
The 1874-2007 dataset does not include the first observed flare, in December 
1859. Nevertheless, \citet{Hod59} reported that the size of the sunspot group 
associated with this event was about $96$ Mm, and one can estimate 
from the drawing of \citet{Car59} that its total area was smaller than 
$6000$ MSH. 

The point marked by a thick $\times$ sign in Fig.~\ref{fig2} is defined 
by the intersection of the $3500$ G and the $6000$ MSH lines. The model 
states that its magnetic flux is $\phi=27\times10^{22}$ Mx. This modeled 
value is much higher than $8\times10^{22}$ Mx, which corresponds both to the 
dominant polarity for the Halloween flares \citep{Kaza10} and to the highest 
flux measured for single active regions, as observed during a sample of 
time-periods between 1998 and 2007 \citep{Parnell09}. The modeled flux for 
this largest sunspot group is nevertheless consistent with the maximum value 
of $20\times10^{22}$ Mx for an active region, as reported by \citet{Zhang10} 
in a very extensive survey, ranging from 1996 to 2008. It remains difficult 
to estimate the highest active region flux which ever occurred. Firstly, 
no magnetic field measurement is available for the April 1947 sunspot 
group. Secondly, the automatic procedure of 
\citet{Zhang10} can lead several active regions to be grouped into 
an apparent single region, while the method of \citet{Parnell09} 
in contrast tends to fragment active region into several pieces. 
For reference, we therefore overplotted both $\phi=8\times10^{22}$ and 
$2\times10^{23}$ Mx values in Fig.~\ref{fig2}. 

The flare energy at the point $\times$, where the magnetic field and 
size of sunspot groups take their extreme values, is $E^{\times}=340\times 
10^{32}$ ergs. This could a priori be considered as the maximum possible 
energy of a solar flare. In addition, it falls within the 
range of stellar superflare energies 
\citep{Mae12}. 
Nevertheless, we argue below that this point is unrealistic for 
observed solar conditions. 

\subsection{Taking into account the fragmentation of flux}

     \begin{figure*}
     \centering
     \includegraphics[width=.84\textwidth]{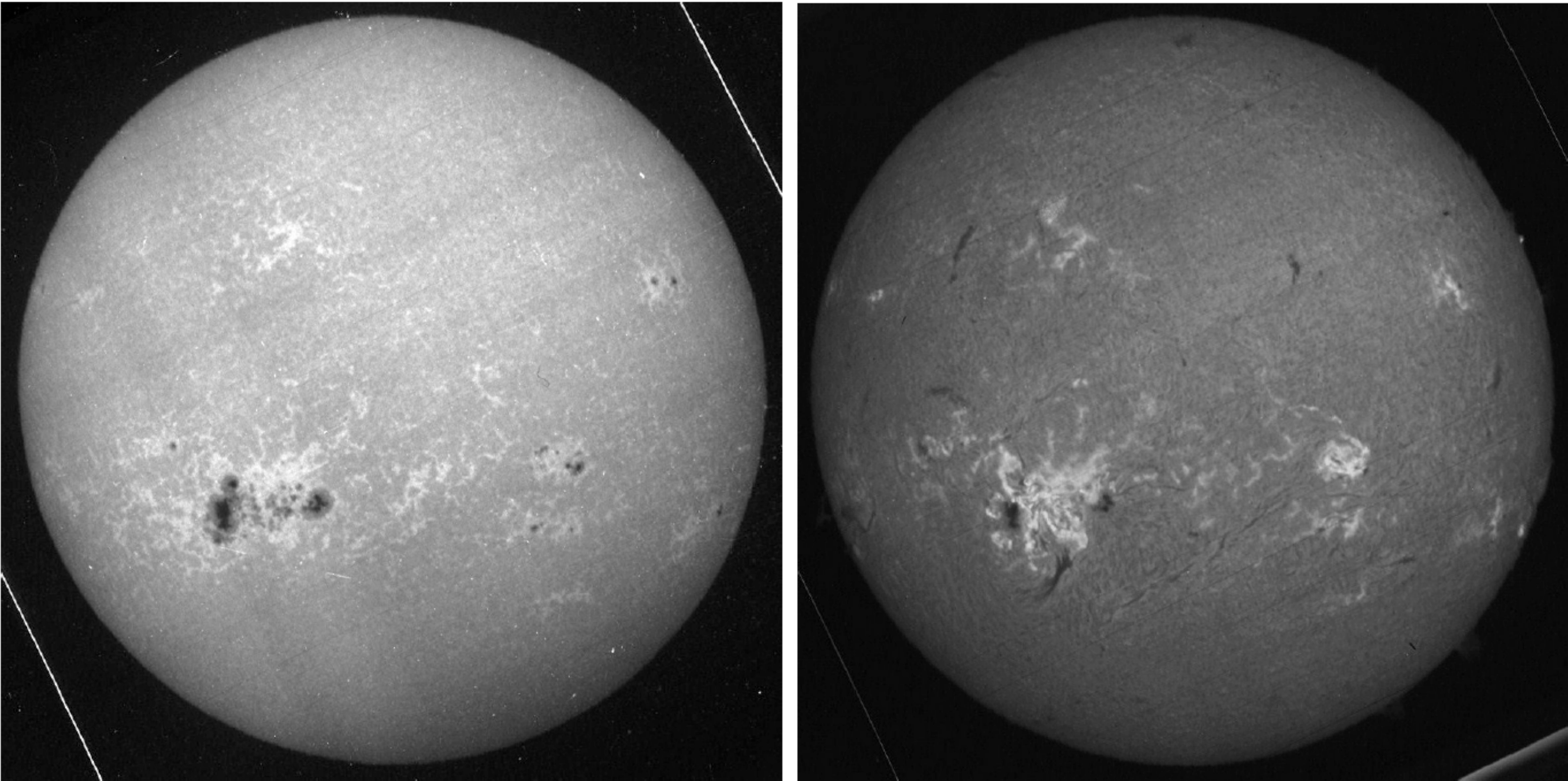}
     \caption{
The largest sunspot group ever reported since the end of the nineteenth 
century, as observed in April 5, 1947 in Ca~{\sc ii} K1v ({\em left}) and H$\alpha$ ({\em 
right}) by the Meudon spectroheliograph. 
           }
     \label{fig3}
     \end{figure*}

All large sunspot groups are highly fragmented, and display many 
episodes of flux emergence and dispersal. We argue that this fragmentation 
is the reason why scaling the model to the whole area of the largest 
sunspot group leads to over estimate the maximum flare energy. 

Firstly, sunspot groups incorporate several big sunspots, ranging 
from a few spots \citep[see e.g.][for February 2011]{Schrijver11} 
to half a dozen \citep[see e.g.][for September 1859 and October 
2003 respectively]{Car59,Schrijver06} and up to more than ten 
\citep[see e.g.][for March 1989 and April 1947 respectively; see 
also Fig.~\ref{fig3}]{Wang91,Nic48}. 
Secondly, these groups typically have a magnetic flux imbalance 
\citep[e.g. $23\%$ for the October 2003 sunspot group][]{Kaza10}, because 
they often emerge within older active regions. This naturally creates 
new magnetic connections to distant regions on the Sun, in addition 
to possibly pre-existing ones. Thirdly, the magnetic shear tends to 
be concentrated along some segments 
only of the polarity inversion lines of a given group \citep{Fal08}. 
This is also true for the April 1947 sunspot group, as evidenced by 
the complex distribution of small filaments (see Fig.~\ref{fig3}). 
This means that a given sunspot group is never energized as a whole. 
These three observational properties are actually consistent with 
the solar convection-driven breaking of large sub-photospheric flux 
tubes into a series of smaller deformed structures, as found in 
numerical simulations \citep{Fan03,JouveSub}. They show 
that these deformed structures should eventually emerge through 
the photosphere as grouped but distinct magnetic bipoles. These 
different bipoles should naturally possess various degrees of 
magnetic shear, and should not be fully magnetically connected 
to each other in the corona. 

So both observational and theoretical arguments suggest that only 
a few sunspots from a whole sunspot group should be involved in 
a given flare. 
%
%
Unfortunately, the fraction of area to be considered, and to be 
compared with the size of the bipole in the model, is difficult 
to estimate. 

We consider the Oct-Nov 2003 flares, for example. Our estimation of 
$L^{\mbox{\rm{\tiny bipole}}} \sim 65$ Mm, as given above, results 
in a modeled sunspot area of $700$ MSH (see Fig.~\ref{fig2}). This 
is about $27\%$ of the maximum area measured for the whole sunspot 
group, which peaked at $2600$ MSH on Oct 31. Another way to 
estimate this fraction is to measure the ratio between the magnetic 
flux swept by the flare ribbons, and that of the whole active 
region. \citet{Qiu07} and \citet{Kaza12} reported a ratio of 
$25\%$ and $31\%$ for the Oct 28 flare, respectively. The same 
authors also reported on a dozen of other events, for which 
one can estimate ratios ranging between $10\%$ and $30\%$, on 
average. 

These considerations lead us to conjecture that $30\%$ at most 
of the area of the largest observed sunspot group, as reported 
by \citet{Nic48} and \citet{Taylor89}, i.e. a maximum of $1800$ MSH, 
can be involved in a flare. This is more than $2.5$ times the area 
of the bipole involved in the Halloween flares. 
In Fig.~\ref{fig2}, we therefore plot another point indicated by a 
thick $+$ sign, located at the intersection of the $3500$ G and the 
$1800$ MSH lines. In the model, this corresponds to $L^{\mbox{\rm{\tiny 
bipole}}}=105$ Mm. The flare energy at this point is $E^{+}=56\times
10^{32}$ ergs. Under the assumptions of the model, and considering 
that it probably corresponds to the most extreme observed solar 
conditions, $E^{+}$ should correspond to the upper limit on solar 
flare energy. 

\subsection{Numerical concerns}

As for all numerical models, various limitations could play a role in changing 
the estimated maximum flare energy $E^{+}$. 

We mentioned above that the simulation did not cover the duration of the full eruption, 
because some numerical instabilities eventually developed. On the one hand, this means 
that our flare energies are slightly under estimated. But on the other hand, 
the low $R_m$ must lead to a weak large-scale diffusion. It should not be 
very strong, however, since the characteristic diffusion time at the scale of 
the modeled bipole can be estimated as $150$ times the duration of the 
simulation. Still, it ought to take away some fraction of the magnetic energy 
released during the simulation, so that our flare energies are slightly 
over estimated. Quantifying the relative importance of both effects is 
unfortunately hard to achieve. 

Moreover, applying different spatial distribution of shear during the pre-flare 
energy storage phase could lead to a different amount of energy release \citep{Fal08}. 
But in our model, the shearing motions were extended all along the polarity 
inversion line in the middle of the flux concentrations. Therefore we argue that 
it will be difficult for different settings to produce significantly higher 
flare energies. 


%
Another concern is that our simulation produces a CME kinetic 
energy which is only $5\%$ of the flare energy. But current observational 
energy estimates imply that the kinetic energy of a CME can be the same 
as \citep{Emslie05} and up to three times higher than \citep{Emslie12} the 
bolometric energy of its associated flare. This strong discrepancy cannot 
be attributed to the fact that our simulation was limited in time. Indeed, 
other 3D (resp. 2.5D) MHD models calculated by independent groups and codes 
predict that no more than $10\%$ (resp. $30\%$) of the total released magnetic 
energy is converted into the CME kinetic energy \citep{Amari03b,Jaco06,Lynch08,Ree10}. 
This means that it is unclear whether the relatively weaker CME kinetic energy 
in our model should be attributed to observational biases, or to numerical 
problems commonly shared by several groups and codes. 

In principle, the validity of the model can also be questioned because 
%
magnetic reconnection is ensured by resistivity, with 
a relatively low magnetic Reynolds number $R_m$ as compared to that of the solar 
corona. This may lead to different reconnection rates from those found in 
collisionless reconnection simulations \citep[see e.g.][]{Aunai11}. The 
reconnection rate is indeed important for the flare energy release in 
fully three-dimensional simulations of solar eruptions. In principle, slower 
(resp. faster) reconnection releases weaker (resp. larger) amounts of 
magnetic energy per unit time. Nevertheless, one might argue that the 
time-integrated energy release, during the whole flare, could be not very 
sensitive to the reconnection rate. However the energy content 
which is available at a given time, within a given pair of pre-reconnecting 
magnetic field lines, strongly depends on how much time these field lines 
have had to stretch ideally \citep[as described in][]{Aula12}, and thus by 
how much their magnetic shear has decreased before they reconnect. 
This explains why the time-evolution of the eruption makes the 
reconnection rate important for time-integrated energy release. 
In our simulation, we measure the reconnection rate from the 
average Mach number $M$ of the reconnection inflown. During the 
eruption, it increases in time from $M\sim0.05$ to $M\sim0.2$ approximately. 
These reconnection rates are fortunately comparable to those obtained 
for collisionless reconnection. 
So we conjecture that the limited physics inside our modeled reconnecting 
current sheet should not have drastic consequences for the flare energies. 
Nevertheless, it should be noted that this result probably does not hold 
for other resistive MHD simulations that use very different $R_m$. 

We foresee that these numerical concerns are probably not extremely 
%
sensitive: the orders of magnitudes that we find for flare energies are likely 
to be correct. But it is difficult at present to assert that we 
estimate flare energies with a precision better than several tens of percents, 
or even more. Therefore we conservatively 
%
round up the upper value $E^{+}$ to $6 \times10^{33}$ ergs. In the future, 
data-driven simulations which can explore the parameter space and which incorporate 
more physics will have to be developed to fine-tune the present analyses. 


\section{Summary and discussion}
\label{secsum}

So as to estimate the maximum possible energy of solar flare, we used 
a dimensionless numerical 3D MHD simulation for solar eruptions 
\citep{Aula10,Aula12}. We had previously shown that this model successfully 
matches the observations of active region magnetic fields, of coronal 
sigmoids, of flare ribbons and loops, of CMEs, and of large-scale propagation 
fronts. 

We scaled the model parameters to physical values. Typical solar active region 
parameters resulted in typically observed magnetic fluxes \citep{Parnell09,Zhang10} 
and flare energies \citep{Kret11,Schrij12}. We then scaled the model using 
the largest measured sunspot magnetic field \citep{Solanki06}, and the area of 
the largest sunspot group ever reported, which developed in March-April 1947 
\citep{Nic48,Taylor89}. 

In addition, we took into account that observations show that 
large sunspots groups are always fragmented into several spots, and are 
never involved in a given flare as a whole. This partitioning can presumably 
be attributed to sub-photospheric convective motions. Since those motions are 
always present because of the solar internal structure \citep{Brun02}, it is difficult 
to imagine that the Sun will ever produce a large sunspot group consisting of a single 
pair of giant sunspots. Based on some approximated geometrical and reconnected 
magnetic flux estimations, we considered that only $30\%$ the area of a given 
sunspot group can be involved in a flare. 

Keeping in mind the assumptions and limitations of the numerical model, these 
scalings resulted in a maximum flare energy of $\sim 6 \times10^{33}$ ergs. 
This is is ten times the energy of the Oct 28, 2003 X17 flare, as reported in 
\citet{Schrij12}. In addition, this value is about six times higher than the maximum 
energy in TSI fluence that can be estimated from the SXR fluence of the 
Nov 4, 2003 X28-40 flare, using the scalings given by \citet{Kret11} and 
\citet{Schrij12}. 
Finally, it lies in the energy range of the weakest superflares that 
were reported by \citet{Mae12} for numerous slowly-rotating and isolated 
Sun-like stars. But it is several orders of magnitude smaller than that 
of strong stellar superflares. 

One could ask what the frequency is at which the Sun can produce a maximum 
flare like this. Observational records since 1874 reveal that the area of 
sunspot groups follow a sharp log-normal distribution \citep{Baumann05,Hat08}. 
Unfortunately, the statistics for sunspot groups larger than $3000$ MSH in 
area are too poor to estimate whether or not this distribution is valid up 
to $6000$ MSH. In addition, neither do all active regions or sunspot groups 
generate flares, nor do they always generate them at the maximum energy, 
as calculated by the model. 
The reason must be that a solar eruption requires a strong magnetic 
sheared polarity inversion line, and current observations show that 
this does not occur in all solar active regions \citep{Fal08}. 
Consequently it is currently difficult to estimate the probability of appearance 
of the strongest flare that we found. We can only refer to 
\citet{Baumann05} and \citet{Hat08}, who reported that the size of sunspot 
groups follows a clear log-normal distribution up to $3000$ MSH, and to 
\citet{Cliv04} and \citet{Schrij12}, who argue that this upper 
limit on flare energy was never reached in any observed solar 
flare since, even including the Carrington event of Sept 1859. 

     \begin{figure}
     \centering
     \includegraphics[width=0.405\textwidth,clip]{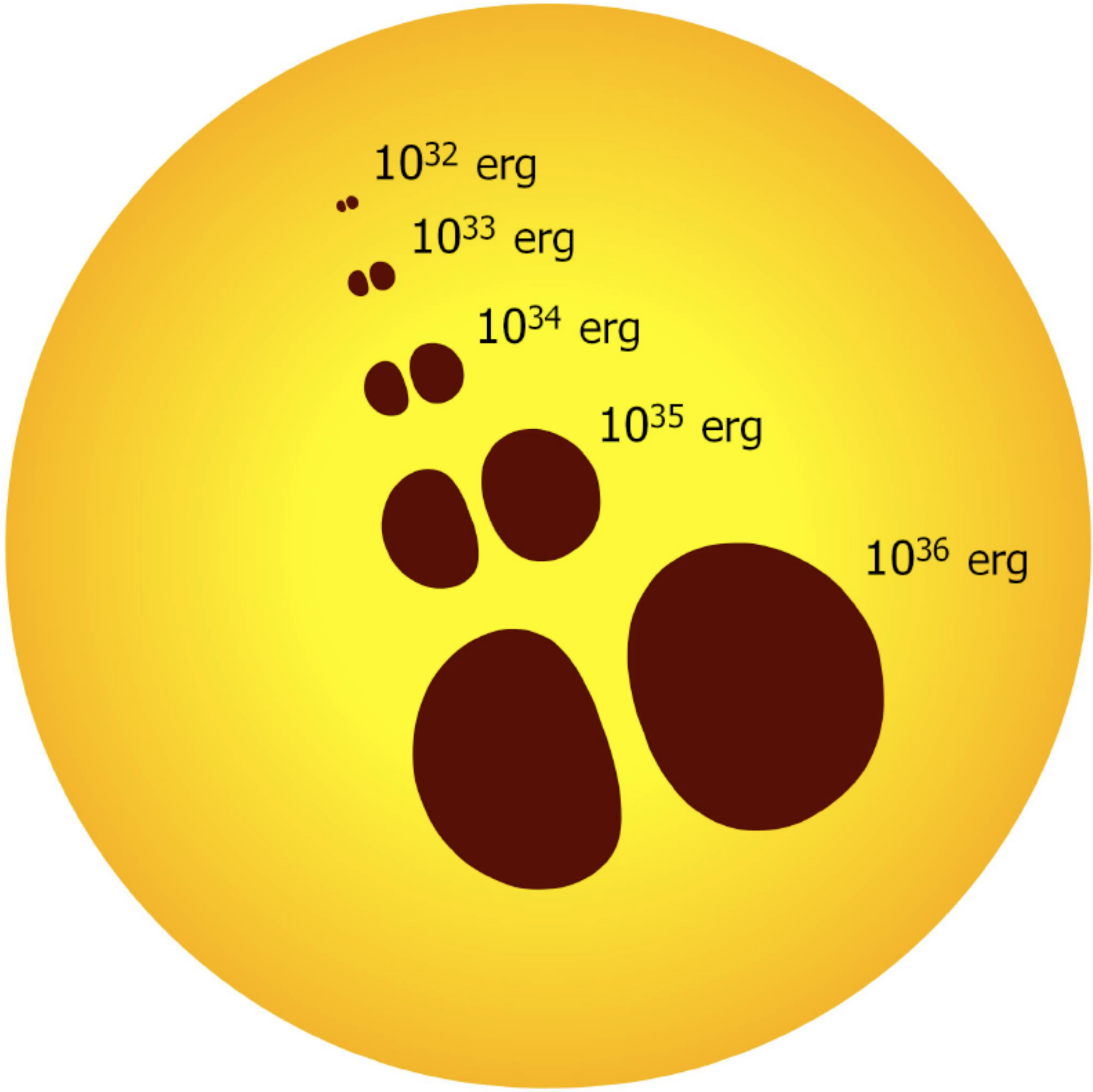}
     \caption{
Schematic representation of several modeled sunspot pairs on the solar disk, with their 
corresponding modeled flare energies. Note that our estimations state that in the 
real Sun, a given pair will often be embedded in a much larger sunspot group, from 
which only the bipole that is shown here will be involved in the flare.
           }
     \label{fig4}
     \end{figure}

When the model is scaled to the strongest measured sunspot magnetic field, 
%
i.e. 3.5~kG, 
%
it can be used to calculate the size of the sunspot pair that is required 
to generate the solar flares 
%
of various energies. We plot those in Fig.~\ref{fig4}. These scalings 
can also be used to relate stellar superflares to starspot sizes. But 
it should be noted that starspot magnetic fields are still difficult to 
measure reliably, and that current estimates put them in the range of 
$2-5$ kG \citep{Ber05}. With these scalings, a 
%
superflare of $10^{36}$ 
ergs requires a very large single pair of spots, whose extent is $48^\circ$ 
in longitude/latitude, at the surface of a Sun-like star. While such spots 
have been observed indirectly in non-Sun-like stars as well as in young 
fast-rotating Sun-like stars 
%
\citep{Ber05,Stras09}, 
%
they have never been reported on the Sun. 

\section{Conclusion}
\label{secccl}

We combined a numerical magnetohydrodynamic model for solar 
eruptions calculated with the OHM code and historical sunspot observations 
starting from the end of the nineteenth century. We concluded 
that the maximum energy of solar flares is about six times 
that of the strongest-ever directly-observed flare of Nov 4, 2003. 

One unaddressed question is whether or not the current solar convective dynamo can 
produce much larger sunspot groups, as required to produce even stronger flares 
according to our results. This seems unlikely, since such giant sunspot groups 
``have not been recorded in four centuries of direct scientific observations 
and in millennia of sunrises and sunsets viewable by anyone around the world'', 
to quote \citet{Schrij12}. 
%
%
It can thus reasonably be assumed that, during the most recent few billion
years while on the main sequence, the Sun never has produced, and never will
produce, a flare more energetic than this upper limit. We thus conjecture
that one condition for Sun-like stars to produce superflares is to host
a dynamo that is much stronger than that of an aged Sun with a rotation
rate exceeding several days.

On the one hand, our results suggest that we have not experienced the largest 
possible solar flare. But on the other hand, and unless the dynamo theory proves 
otherwise, our results also provide an upper limit for extreme space weather 
conditions, that does not exceed those related to past observed flares by much. 


\begin{acknowledgements}
The MHD calculations were done on the quadri-core bi-Xeon computers of the Cluster of 
the Division Informatique de l'Observatoire de Paris (DIO). The historical Meudon 
spectroheliograph observations were digitalized by I.~Bual\'e, and are available in 
the BASS2000 database. The work of MJ is funded by a contract from the AXA Research Fund.  
\end{acknowledgements}


\bibliographystyle{aa} 
\bibliography{aulanierREF}  

\begin{thebibliography}{68}
\expandafter\ifx\csname natexlab\endcsname\relax\def\natexlab#1{#1}\fi

\bibitem[{{Amari} {et~al.}(2003){Amari}, {Luciani}, {Aly}, {Mikic}, \&
  {Linker}}]{Amari03b}
{Amari}, T., {Luciani}, J.~F., {Aly}, J.~J., {Mikic}, Z., \& {Linker}, J. 2003,
  \apj, 595, 1231

\bibitem[{{Archontis} {et~al.}(2004){Archontis}, {Moreno-Insertis},
  {Galsgaard}, {Hood}, \& {O'Shea}}]{Arch04}
{Archontis}, V., {Moreno-Insertis}, F., {Galsgaard}, K., {Hood}, A., \&
  {O'Shea}, E. 2004, \aap, 426, 1047

\bibitem[{{Aulanier} {et~al.}(2005){Aulanier}, {D{\'e}moulin}, \&
  {Grappin}}]{Aula05a}
{Aulanier}, G., {D{\'e}moulin}, P., \& {Grappin}, R. 2005, \aap, 430, 1067

\bibitem[{{Aulanier} {et~al.}(2012){Aulanier}, {Janvier}, \&
  {Schmieder}}]{Aula12}
{Aulanier}, G., {Janvier}, M., \& {Schmieder}, B. 2012, \aap, 543, A110

\bibitem[{{Aulanier} {et~al.}(2010){Aulanier}, {T{\"o}r{\"o}k}, {D{\'e}moulin},
  \& {DeLuca}}]{Aula10}
{Aulanier}, G., {T{\"o}r{\"o}k}, T., {D{\'e}moulin}, P., \& {DeLuca}, E.~E.
  2010, \apj, 708, 314

\bibitem[{{Aunai} {et~al.}(2011){Aunai}, {Belmont}, \& {Smets}}]{Aunai11}
{Aunai}, N., {Belmont}, G., \& {Smets}, R. 2011, Journal of Geophysical
  Research (Space Physics), 116, 9232

\bibitem[{{Baumann} \& {Solanki}(2005)}]{Baumann05}
{Baumann}, I. \& {Solanki}, S.~K. 2005, \aap, 443, 1061

\bibitem[{{Berdyugina}(2005)}]{Ber05}
{Berdyugina}, S.~V. 2005, Living Reviews in Solar Physics, 2, 8

\bibitem[{{Brodrick} {et~al.}(2005){Brodrick}, {Tingay}, \&
  {Wieringa}}]{Brod05}
{Brodrick}, D., {Tingay}, S., \& {Wieringa}, M. 2005, Journal of Geophysical
  Research (Space Physics), 110, 9

\bibitem[{{Brun} \& {Toomre}(2002)}]{Brun02}
{Brun}, A.~S. \& {Toomre}, J. 2002, \apj, 570, 865

\bibitem[{{Carrington}(1859)}]{Car59}
{Carrington}, R.~C. 1859, \mnras, 20, 13

\bibitem[{{Cliver} \& {Svalgaard}(2004)}]{Cliv04}
{Cliver}, E.~W. \& {Svalgaard}, L. 2004, \solphys, 224, 407

\bibitem[{{Delann{\'e}e} {et~al.}(2008){Delann{\'e}e}, {T{\"o}r{\"o}k},
  {Aulanier}, \& {Hochedez}}]{Dela08}
{Delann{\'e}e}, C., {T{\"o}r{\"o}k}, T., {Aulanier}, G., \& {Hochedez}, J.-F.
  2008, \solphys, 247, 123

\bibitem[{{D{\'e}moulin} \& {Aulanier}(2010)}]{DemAula10}
{D{\'e}moulin}, P. \& {Aulanier}, G. 2010, \apj, 718, 1388

\bibitem[{{Emslie} {et~al.}(2005){Emslie}, {Dennis}, {Holman}, \&
  {Hudson}}]{Emslie05}
{Emslie}, A.~G., {Dennis}, B.~R., {Holman}, G.~D., \& {Hudson}, H.~S. 2005,
  Journal of Geophysical Research (Space Physics), 110, 11103

\bibitem[{{Emslie} {et~al.}(2012){Emslie}, {Dennis}, {Shih}, {Chamberlin},
  {Mewaldt}, {Moore}, {Share}, {Vourlidas}, \& {Welsch}}]{Emslie12}
{Emslie}, A.~G., {Dennis}, B.~R., {Shih}, A.~Y., {et~al.} 2012, \apj, 759, 71

\bibitem[{{Falconer} {et~al.}(2008){Falconer}, {Moore}, \& {Gary}}]{Fal08}
{Falconer}, D.~A., {Moore}, R.~L., \& {Gary}, G.~A. 2008, \apj, 689, 1433

\bibitem[{{Fan} {et~al.}(2003){Fan}, {Abbett}, \& {Fisher}}]{Fan03}
{Fan}, Y., {Abbett}, W.~P., \& {Fisher}, G.~H. 2003, \apj, 582, 1206

\bibitem[{{Forbes} {et~al.}(2006){Forbes}, {Linker}, {Chen}, {Cid}, {K{\'o}ta},
  {Lee}, {Mann}, {Miki{\'c}}, {Potgieter}, {Schmidt}, {Siscoe}, {Vainio},
  {Antiochos}, \& {Riley}}]{Forbes06}
{Forbes}, T.~G., {Linker}, J.~A., {Chen}, J., {et~al.} 2006, Space Science
  Reviews, 123, 251

\bibitem[{{Green} {et~al.}(2011){Green}, {Kliem}, \& {Wallace}}]{Green11}
{Green}, L.~M., {Kliem}, B., \& {Wallace}, A.~J. 2011, \aap, 526, A2

\bibitem[{{Hathaway} \& {Choudhary}(2008)}]{Hat08}
{Hathaway}, D.~H. \& {Choudhary}, D.~P. 2008, \solphys, 250, 269

\bibitem[{{Hodgson}(1859)}]{Hod59}
{Hodgson}, R. 1859, \mnras, 20, 15

\bibitem[{{Jacobs} {et~al.}(2006){Jacobs}, {Poedts}, \& {van der
  Holst}}]{Jaco06}
{Jacobs}, C., {Poedts}, S., \& {van der Holst}, B. 2006, \aap, 450, 793

\bibitem[{{Jouve} {et~al.}(2013){Jouve}, {Brun}, \& {Aulanier}}]{JouveSub}
{Jouve}, L., {Brun}, A.~S., \& {Aulanier}, G. 2013, \apj, in press

\bibitem[{{Kazachenko} {et~al.}(2010){Kazachenko}, {Canfield}, {Longcope}, \&
  {Qiu}}]{Kaza10}
{Kazachenko}, M.~D., {Canfield}, R.~C., {Longcope}, D.~W., \& {Qiu}, J. 2010,
  \apj, 722, 1539

\bibitem[{{Kazachenko} {et~al.}(2012){Kazachenko}, {Canfield}, {Longcope}, \&
  {Qiu}}]{Kaza12}
{Kazachenko}, M.~D., {Canfield}, R.~C., {Longcope}, D.~W., \& {Qiu}, J. 2012,
  \solphys, 277, 165

\bibitem[{{Kliem} \& {T{\"o}r{\"o}k}(2006)}]{KliTor06}
{Kliem}, B. \& {T{\"o}r{\"o}k}, T. 2006, Physical Review Letters, 96, 255002

\bibitem[{{Kretzschmar}(2011)}]{Kret11}
{Kretzschmar}, M. 2011, \aap, 530, A84

\bibitem[{{Lin} \& {Forbes}(2000)}]{LinForbes00}
{Lin}, J. \& {Forbes}, T.~G. 2000, \jgr, 105, 2375

\bibitem[{{Livingston} {et~al.}(2012){Livingston}, {Penn}, \&
  {Svalgaard}}]{Living12}
{Livingston}, W., {Penn}, M.~J., \& {Svalgaard}, L. 2012, \apjl, 757, L8

\bibitem[{{Lynch} {et~al.}(2008){Lynch}, {Antiochos}, {DeVore}, {Luhmann}, \&
  {Zurbuchen}}]{Lynch08}
{Lynch}, B.~J., {Antiochos}, S.~K., {DeVore}, C.~R., {Luhmann}, J.~G., \&
  {Zurbuchen}, T.~H. 2008, \apj, 683, 1192

\bibitem[{{Maehara} {et~al.}(2012){Maehara}, {Shibayama}, {Notsu}, {Notsu},
  {Nagao}, {Kusaba}, {Honda}, {Nogami}, \& {Shibata}}]{Mae12}
{Maehara}, H., {Shibayama}, T., {Notsu}, S., {et~al.} 2012, \nat, 485, 478

\bibitem[{{Masson} {et~al.}(2009){Masson}, {Klein}, {B{\"u}tikofer},
  {Fl{\"u}ckiger}, {Kurt}, {Yushkov}, \& {Krucker}}]{MassonKlein09}
{Masson}, S., {Klein}, K.-L., {B{\"u}tikofer}, R., {et~al.} 2009, \solphys,
  257, 305

\bibitem[{{McCracken} {et~al.}(2001){McCracken}, {Dreschhoff}, {Zeller},
  {Smart}, \& {Shea}}]{Mc01}
{McCracken}, K.~G., {Dreschhoff}, G.~A.~M., {Zeller}, E.~J., {Smart}, D.~F., \&
  {Shea}, M.~A. 2001, \jgr, 106, 21585

\bibitem[{{Moore} {et~al.}(2012){Moore}, {Falconer}, \& {Sterling}}]{Moore12}
{Moore}, R.~L., {Falconer}, D.~A., \& {Sterling}, A.~C. 2012, \apj, 750, 24

\bibitem[{{Moore} {et~al.}(2001){Moore}, {Sterling}, {Hudson}, \&
  {Lemen}}]{Moore01}
{Moore}, R.~L., {Sterling}, A.~C., {Hudson}, H.~S., \& {Lemen}, J.~R. 2001,
  \apj, 552, 833

\bibitem[{{Nakariakov} {et~al.}(1999){Nakariakov}, {Ofman}, {Deluca},
  {Roberts}, \& {Davila}}]{Nak99}
{Nakariakov}, V.~M., {Ofman}, L., {Deluca}, E.~E., {Roberts}, B., \& {Davila},
  J.~M. 1999, Science, 285, 862

\bibitem[{{Nicholson}(1948)}]{Nic48}
{Nicholson}, S.~B. 1948, \pasp, 60, 98

\bibitem[{{Parnell} {et~al.}(2009){Parnell}, {DeForest}, {Hagenaar},
  {Johnston}, {Lamb}, \& {Welsch}}]{Parnell09}
{Parnell}, C.~E., {DeForest}, C.~E., {Hagenaar}, H.~J., {et~al.} 2009, \apj,
  698, 75

\bibitem[{{Pevtsov} {et~al.}(2011){Pevtsov}, {Nagovitsyn}, {Tlatov}, \&
  {Rybak}}]{Pevtsov11}
{Pevtsov}, A.~A., {Nagovitsyn}, Y.~A., {Tlatov}, A.~G., \& {Rybak}, A.~L. 2011,
  \apjl, 742, L36

\bibitem[{{Priest} \& {Forbes}(2002)}]{Priest02}
{Priest}, E.~R. \& {Forbes}, T.~G. 2002, \aapr, 10, 313

\bibitem[{{Pulkkinen}(2007)}]{Pulk07}
{Pulkkinen}, T. 2007, Living Reviews in Solar Physics, 4, 1

\bibitem[{{Qiu} {et~al.}(2007){Qiu}, {Hu}, {Howard}, \& {Yurchyshyn}}]{Qiu07}
{Qiu}, J., {Hu}, Q., {Howard}, T.~A., \& {Yurchyshyn}, V.~B. 2007, \apj, 659,
  758

\bibitem[{{Reeves} {et~al.}(2010){Reeves}, {Linker}, {Miki{\'c}}, \&
  {Forbes}}]{Ree10}
{Reeves}, K.~K., {Linker}, J.~A., {Miki{\'c}}, Z., \& {Forbes}, T.~G. 2010,
  \apj, 721, 1547

\bibitem[{{Savcheva} {et~al.}(2012){Savcheva}, {Pariat}, {van Ballegooijen},
  {Aulanier}, \& {DeLuca}}]{Sav12}
{Savcheva}, A., {Pariat}, E., {van Ballegooijen}, A., {Aulanier}, G., \&
  {DeLuca}, E. 2012, \apj, 750, 15

\bibitem[{{Schaefer} {et~al.}(2000){Schaefer}, {King}, \&
  {Deliyannis}}]{Schae00}
{Schaefer}, B.~E., {King}, J.~R., \& {Deliyannis}, C.~P. 2000, \apj, 529, 1026

\bibitem[{{Schmieder} {et~al.}(1987){Schmieder}, {Forbes}, {Malherbe}, \&
  {Machado}}]{Sch87}
{Schmieder}, B., {Forbes}, T.~G., {Malherbe}, J.~M., \& {Machado}, M.~E. 1987,
  \apj, 317, 956

\bibitem[{{Schrijver}(2009)}]{SchrijverCospar09}
{Schrijver}, C.~J. 2009, Advances in Space Research, 43, 739

\bibitem[{{Schrijver} {et~al.}(2011){Schrijver}, {Aulanier}, {Title}, {Pariat},
  \& {Delann{\'e}e}}]{Schrijver11}
{Schrijver}, C.~J., {Aulanier}, G., {Title}, A.~M., {Pariat}, E., \&
  {Delann{\'e}e}, C. 2011, \apj, 738, 167

\bibitem[{{Schrijver} {et~al.}(2012){Schrijver}, {Beer}, {Baltensperger},
  {Cliver}, {G{\"u}del}, {Hudson}, {McCracken}, {Osten}, {Peter}, {Soderblom},
  {Usoskin}, \& {Wolff}}]{Schrij12}
{Schrijver}, C.~J., {Beer}, J., {Baltensperger}, U., {et~al.} 2012, Journal of
  Geophysical Research (Space Physics), 117, 8103

\bibitem[{{Schrijver} {et~al.}(2006){Schrijver}, {Hudson}, {Murphy}, {Share},
  \& {Tarbell}}]{Schrijver06}
{Schrijver}, C.~J., {Hudson}, H.~S., {Murphy}, R.~J., {Share}, G.~H., \&
  {Tarbell}, T.~D. 2006, \apj, 650, 1184

\bibitem[{{Schrijver} \& {Title}(2011)}]{SchrijverTitle11}
{Schrijver}, C.~J. \& {Title}, A.~M. 2011, Journal of Geophysical Research
  (Space Physics), 116, 4108

\bibitem[{{Schwenn}(2006)}]{Schwenn06}
{Schwenn}, R. 2006, Living Reviews in Solar Physics, 3, 2

\bibitem[{{Shibata} {et~al.}(1995){Shibata}, {Masuda}, {Shimojo}, {Hara},
  {Yokoyama}, {Tsuneta}, {Kosugi}, \& {Ogawara}}]{Shibata95}
{Shibata}, K., {Masuda}, S., {Shimojo}, M., {et~al.} 1995, \apjl, 451, L83

\bibitem[{{Solanki} {et~al.}(2006){Solanki}, {Inhester}, \&
  {Sch{\"u}ssler}}]{Solanki06}
{Solanki}, S.~K., {Inhester}, B., \& {Sch{\"u}ssler}, M. 2006, Reports on
  Progress in Physics, 69, 563

\bibitem[{{Strassmeier}(2009)}]{Stras09}
{Strassmeier}, K.~G. 2009, \aapr, 17, 251

\bibitem[{{Taylor}(1989)}]{Taylor89}
{Taylor}, P.~O. 1989, Journal of the American Association of Variable Star
  Observers (JAAVSO), 18, 65

\bibitem[{{Tsurutani} {et~al.}(2003){Tsurutani}, {Gonzalez}, {Lakhina}, \&
  {Alex}}]{Tsu03}
{Tsurutani}, B.~T., {Gonzalez}, W.~D., {Lakhina}, G.~S., \& {Alex}, S. 2003,
  Journal of Geophysical Research (Space Physics), 108, 1268

\bibitem[{{van Ballegooijen} \& {Martens}(1989)}]{vanBalle89}
{van Ballegooijen}, A.~A. \& {Martens}, P.~C.~H. 1989, \apj, 343, 971

\bibitem[{{van Driel-Gesztelyi} {et~al.}(2003){van Driel-Gesztelyi},
  {D{\'e}moulin}, {Mandrini}, {Harra}, \& {Klimchuk}}]{vanDriel03}
{van Driel-Gesztelyi}, L., {D{\'e}moulin}, P., {Mandrini}, C.~H., {Harra}, L.,
  \& {Klimchuk}, J.~A. 2003, \apj, 586, 579

\bibitem[{{Vourlidas} {et~al.}(2010){Vourlidas}, {Howard}, {Esfandiari},
  {Patsourakos}, {Yashiro}, \& {Michalek}}]{Vour10}
{Vourlidas}, A., {Howard}, R.~A., {Esfandiari}, E., {et~al.} 2010, \apj, 722,
  1522

\bibitem[{{Wang} {et~al.}(1991){Wang}, {Tang}, {Zirin}, \& {Ai}}]{Wang91}
{Wang}, H., {Tang}, F., {Zirin}, H., \& {Ai}, G. 1991, \apj, 380, 282

\bibitem[{{Warren} {et~al.}(2011){Warren}, {O'Brien}, \& {Sheeley}}]{Warren11}
{Warren}, H.~P., {O'Brien}, C.~M., \& {Sheeley}, Jr., N.~R. 2011, \apj, 742, 92

\bibitem[{{Willis} \& {Stephenson}(2001)}]{Willis01}
{Willis}, D.~M. \& {Stephenson}, F.~R. 2001, Annales Geophysicae, 19, 289

\bibitem[{{Wolff} {et~al.}(2012){Wolff}, {Bigler}, {Curran}, {Dibb}, {Frey},
  {Legrand}, \& {McConnell}}]{Wolff12}
{Wolff}, E.~W., {Bigler}, M., {Curran}, M.~A.~J., {et~al.} 2012, \grl, 39, 8503

\bibitem[{{Woods} {et~al.}(2004){Woods}, {Eparvier}, {Fontenla}, {Harder},
  {Kopp}, {McClintock}, {Rottman}, {Smiley}, \& {Snow}}]{Woods04}
{Woods}, T.~N., {Eparvier}, F.~G., {Fontenla}, J., {et~al.} 2004, \grl, 31,
  10802

\bibitem[{{Zhang} {et~al.}(2010){Zhang}, {Wang}, \& {Liu}}]{Zhang10}
{Zhang}, J., {Wang}, Y., \& {Liu}, Y. 2010, \apj, 723, 1006

\bibitem[{{Zharkov} {et~al.}(2011){Zharkov}, {Green}, {Matthews}, \&
  {Zharkova}}]{Zhar11}
{Zharkov}, S., {Green}, L.~M., {Matthews}, S.~A., \& {Zharkova}, V.~V. 2011,
  \apjl, 741, L35

\end{thebibliography}
\IfFileExists{\jobname.bbl}{} {\typeout{}
\typeout{***************************************************************}
\typeout{***************************************************************}
\typeout{** Please run "bibtex \jobname" to obtain the bibliography} 
\typeout{** and re-run "latex \jobname" twice to fix references} 
\typeout{***************************************************************}
\typeout{***************************************************************}
\typeout{}}

\end{document}